# Fundamental limit of the microresonator field uniformity and slow light enabled angstrom-precise straight-line translation

## M. Sumetsky

*Aston Institute of Photonic Technologies, Aston University, Birmingham B4 7ET, UK*
m.sumetsky@aston.ac.uk

We determine the fundamental limit of the microresonator field uniformity. It can be achieved in a specially designed microresonator, called a bat microresonator, fabricated at the optical fiber surface. We show that the relative nonuniformity of an eigenmode amplitude along the axial length $L$ of an ideal bat microresonator cannot be smaller than $\frac{1}{3}\pi^2 n_r^4 \lambda^{-4} Q^{-2} L^4$, where $n_r$, $\lambda$ and $Q$ are its refractive index, the eigenmode wavelength and Q-factor. In the absence of losses ($Q = \infty$), this eigenmode has the amplitude independent of axial coordinate and zero axial speed (i.e., is stopped) within the length $L$. For a silica microresonator with $Q = 10^8$ this eigenmode has the axial speed ~ $10^{-4}c$, where $c$ is the speed of light in vacuum, and its nonuniformity along the length $L = 100$ μm at wavelength $\lambda = 1.5$ μm is ~ $10^{-7}$. For a realistic fiber with diameter 100 μm and surface roughness 0.2 nm, the smallest eigenmode nonuniformity is ~ 0.0003. As an application, we consider a bat microresonator evanescently coupled to high Q-factor silica microspheres which serves as a reference supporting the angstrom-precise straight-line translation over the distance $L$ exceeding a hundred microns.

A monochromatic optical field in an ideal unconfined uniform medium with refractive index $n_r$ can have the form of a plane wave $E = A\exp(ikx)$ with propagation constant $k = 2\pi n_r/\lambda$ and wavelength $\lambda$. The amplitude of this field, $|E| = A$, is *independent of coordinates*. The question if the amplitude of a confined optical field – e.g., an eigenmode of an optical resonator – can be *uniform in a finite spatial area* is less intuitive: this field is localized due to reflections which cause interference and spatial variation of its amplitude. For example, light confined in one dimension along axis $x$ oscillates, roughly, as $A\cos(kx)$ and the condition of its constant amplitude requires $k = 2\pi n_r/\lambda = 0$. Thus, for finite $n_r$, the amplitude of a confined field cannot be spatially uniform in 1D unless its frequency $\nu = c/\lambda$ tends to zero, i.e., unless the field is a stationary electric field.

The situation in 2D and 3D is different since then the field can rotate along a closed path with a finite propagation constant without reflections and, simultaneously, can have zero propagation constant along the direction transverse to this path. As an example, consider a SNAP bottle microresonator (BMR) introduced along the surface of an optical fiber [1]. Whispering gallery mode (WGM) eigenstates of such resonator with azimuthal, radial, and axial quantum numbers $m$, $p$, and $q$ are determined in cylindrical coordinates $(z, \rho, \varphi)$, as $E_{mpq}^{(BMR)}(z, \rho, \varphi) = e^{im\varphi} F_{mp}(\rho) \Psi_{mpq}(z)$. The eigenwavelengths $\lambda_{mpq}$ of this BMR are the eigenvalues of the one-dimensional wave equation for $\Psi_{mpq}(z)$ [1]:

$$\frac{d^2 \Psi_{mpq}(z)}{dz^2} + \beta_{mpq}^2(z)\Psi_{mpq}(z) = 0, \quad (1)$$
$$\beta_{mpq}(z) = 2^{3/2} \pi n_r (\lambda_{mp}^{(c)})^{-3/2} (\lambda_{mp}(z) - \lambda_{mpq} + i\gamma_{mp})^{1/2}.$$

Here $\gamma_{mp}$ determines the WGM propagation loss, $\lambda_{mp}^{(c)}(z)$ is the fiber cutoff wavelength depending on the axial coordinate $z$, and $n_r$ is the effective refractive index of the fiber. For a fiber having refractive index distribution $n_r(\rho)$, function $F_{m,p}(\rho)$ satisfies the equation [2]:

$$\frac{d^2 F_{mp}}{d\rho^2} + \frac{1}{\rho}\frac{dF_{mp}}{d\rho} + \left(k_0^2 n_r^2(\rho) - \beta^2 - \frac{m^2}{\rho^2}\right)F_{mp} = 0. \quad (2)$$

In Ref. [3] we presented an analytical example of a BMR having an eigenmode with $q = 0$ (fundamental axial eigenmode) which amplitude is uniform along a fraction of the resonator length $z_1 < z < z_2$. It was shown that the BMR should be uniform along this length and have "ears" at edges (Fig. 1(a)). Since the profile of this BMR resembles the profile of a bat, this BMR was called the bat microresonator (BatMR). It is straightforward to generalize results of Ref. [3] and design a BatMR having an eigenmode with arbitrary $q$ which amplitude is uniform along the fiber segment $z_1 < z < z_2$. Below, we call this eigenmode the *bat mode*. For larger quantum numbers $q$, the ears of BatMRs should be appropriately larger. As an example, Figs. 1(a)-(c) show the nanoscale profile, variation of cutoff wavelength $\lambda_{mp}^{(c)}(z)$, and the bat mode profile determined from Eq. (1) for $q = 2$. We constructed the profile of $\lambda_{mp}^{(c)}(z)$ shown in Fig. 1(b) numerically by varying the size of ears at the edges of the segment $z_1 < z < z_2$ until the eigenwavelength $\lambda_{mp2}$ becomes equal to the value of $\lambda_{mp}^{(c)}(z)$ at this segment. Experimentally, a BatMR can be created using the SNAP technology [1, 4, 5, 6, 7] as follows. First, a BMR with sufficiently large slopes at the edges and uniform center part (e.g., a rectangular BMR [6]) is formed. Next, the ears at the BMR edges can be introduced by iterations with CO$_2$ laser beam shots [5], so that $\lambda_{mpq}$ rises and, finally, coincides with $\lambda_{mp}^{(c)}(z_c)$.

In the ideal case of no losses, $\gamma_{mp} = 0$, a BatMR can have a series of bat modes corresponding to different azimuthal quantum numbers $m$, which amplitude is independent of axial coordinate and has zero axial speed (i.e., is stopped) along the segment $z_1 < z < z_2$. For a lossy BatMR with an ideally uniform section along the segment $z_1 < z < z_2$, the best possible WGM uniformity is determined by the value of losses $\gamma_{mp} = \lambda_{mp}/Q$, where $Q$ is the BatMR Q-factor. From Eq. (1), the smallest possible variation of an eigenmode $\Psi_{mpq}(z)$ is achieved for smallest possible propagation constant $\beta_{mpq}(z) \equiv \beta_{mpq}^{(0)} = 2^{1/2}(i+1)\pi n_r (\lambda_{mp}^{(c)})^{-3/2}(\gamma_{mp})^{1/2}$, i.e., when the cutoff wavelength $\lambda_{mp}^{(c)}(z)$ is constant and equal to the eigenwavelength of this eigenmode, $\lambda_{mp}^{(c)}(z) \equiv \lambda_{mpq}$ (Figs. 1(b) and (c)). The axial speed of this bat mode, at the segment $z_1 < z < z_2$ is $V_{mpq} = (2\pi)^{-1} c \lambda_{mp} \text{Re}(\beta_{mpq}^{(0)}) = (2Q)^{-1/2} n_r c$, where $c$ is the speed of light. For a silica microresonator with $n_r = 1.44$ and $Q = 10^8$ this speed is $\sim c/10000$. Using Eq. (1) and the approximate expression for the bat mode radial dependence (see, e.g., [8]), we determine the evanescent WGM inside this segment assuming that it is *symmetric* with respect to the segment center $z_C = (z_1 + z_2)/2$ as

$$E_{mpq}^{(BMR)}(z,\rho,\varphi) \sim \exp(im\varphi)\exp(-\zeta_{mp}(\rho - r_0))\cos(\beta_{mpq}^{(0)}(z - z_C)),$$
$$\beta_{mpq}^{(0)} = 2\pi n_r(\lambda_{mp}^{(c)})^{-3/2}\gamma_{mp}^{1/2}(1+i), \quad \zeta_{mp} = 2\pi(n_r^2 - 1)^{1/2}(\lambda_{mp}^{(c)})^{-1}. \quad \textbf{(3)}$$

From Eq. (3), in a close vicinity of the BatMR surface, when $\zeta_{mpq}(\rho - r_0) \ll 1$, and $|\beta_{mpq}^{(0)}(z - z_C)| \ll 1$, we find the relative variation of the amplitude of the evanescent field along the segment $z_1 < z < z_2$ as

$$\varepsilon(z) = |\Delta E_{mpq}^{(BMR)}(z,r_0,\varphi) / E_{mpq}^{(BMR)}(z,r_0,\varphi)| = \frac{16\pi^4 n_r^4 (z - z_C)^4}{3Q^2(\lambda_{mp}^{(c)})^4}, \quad \textbf{(4)}$$

while the field amplitude is constant at the surface

$$\Delta\rho(z) = \rho(z) - r_0 = \frac{8\pi^3 n_r^4 (z - z_C)^4}{3Q^2(n_r^2 - 1)^{1/2}(\lambda_{mp}^{(c)})^3}. \quad \textbf{(5)}$$

From Eq. (4), we find the smallest possible field uniformity along the length $L$ of the BatMR surface as $\varepsilon\left(\frac{L}{2}\right) = \frac{1}{3}\pi^2 n_r^4 \lambda^{-4} Q^{-2} L^4$. Fig. 1(g) shows the profiles of $\varepsilon(z)$ and $\Delta\rho(z)$ for a silica BMR with $n_r = 1.44$ at wavelength $\lambda_{mpq} = 1.55$ μm for different Q-factors, $Q = 10^9, 10^8, 10^7,$ and $10^6$.

Generally, it is possible to design a BMR having an eigenmode which amplitude is uniform within a fraction of its *surface, cross-section, and volume*. We assume, as previously, that at the cutoff wavelength $\lambda = 2\pi/k_0 = \lambda_{mp}^{(c)}$ the propagation constant of this mode $\beta = 0$ and this WGM is uniform along the BMR segment $z_1 < z < z_2$. Provided that $\beta = 0$, it follows from Eq. (2) that oscillations along a radial segment $\rho_1 < \rho < \rho_2$ can be suppressed if the squared effective refractive index, $n_{eff}^2(\rho) = n_r^2(\rho) - m^2/(k_0\rho)^2$, is zero at this segment (see Supplementary Material in [9] where $\beta_{mpq}$ for radially nonuniform $n_r(\rho)$ was determined). Analogous to the design shown in Fig. 1(a)-(c), we introduce two "ears" at the edges of this segment (Fig. 1(d) and (e)) and adjust their sizes to ensure $|E_{mp}(z,\rho,\varphi)| = |F_{mp}(\rho)| = const$ for $\rho_1 < \rho < \rho_2$ (Fig. 1(f)). To minimize the required modification of the refractive index and ensure the condition of total internal reflection, this radial segment should be sufficiently short and close to the fiber surface.

Besides the fundamental limit of the microresonator field uniformity caused by optical losses, the uniformity of BMR field $E_{mpq}^{(BMR)}(z,\rho,\varphi)$ is determined by the uniformity of the fiber. Here we are interested in the nanoscale effective radius variation (ERV) of an optical fiber $\Delta r_{eff}(z)$ which is determined through the profile of its cutoff wavelength variation $\Delta\lambda_{mp}^{(c)}(z)$ as $\Delta r_{eff}(z) = r_0 \Delta\lambda_{mp}^{(c)}(z)/\lambda_{mp}^{(c)}$, where $r_0$ is fiber radius. The predetermined $\Delta r_{eff}(z)$ and $\Delta\lambda_{mp}^{(c)}(z)$ can be introduced with subangstrom precision using fabrication methods of the SNAP technology (see e.g., [1, 3 - 7]). The possible measurement precision of $\Delta\lambda_{mp}^{(c)}(z)$ is determined by the fiber and microresonator Q-factors. Assuming $Q = 10^8$ [10], $r_0 = 50$ μm and $\lambda_{mp}^{(c)} = 1.5$ μm, we find that $\Delta\lambda_{mp}^{(c)}(z)$ can be measured with the precision better than $\lambda_{mp}^{(c)}(z)/Q = 0.01$ pm and the ERV $\Delta r_{eff}(z)$ can be measured with the precision better than 0.3 pm. The ERV experimentally measured in [11] was less than 0.2 nm over the axial lengths of hundreds of microns.

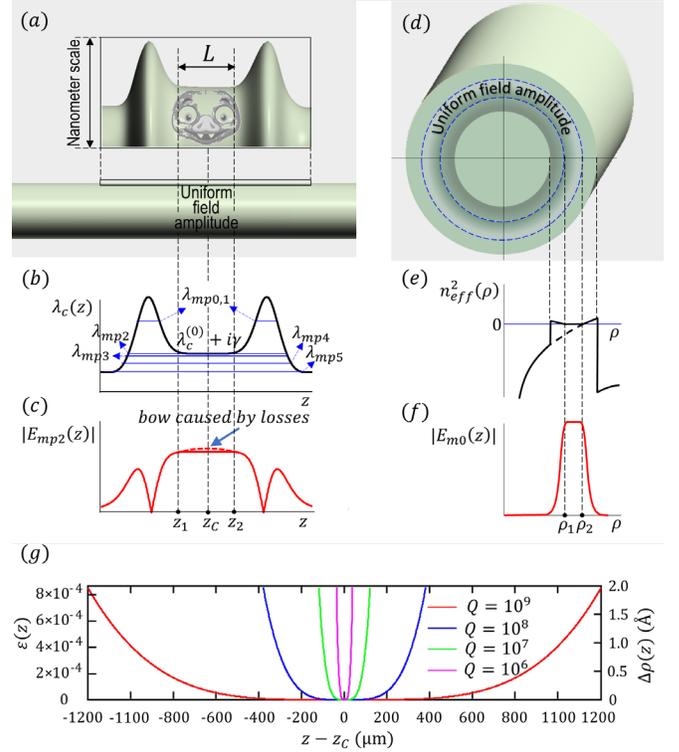

Fig. 1. (a), (d) Illustration of an optical fiber with (b) cutoff wavelength distribution (solid black curve) and (e) original (dashed black curve) and modified (solid black curve) cross-sectional effective refractive index squared distribution designed to have a portion with (c) uniform axial distribution and (f) uniform radial distribution of the WGM amplitude. Notice asymmetry of "ears" of $n_{eff}^2(\rho)$ in (e) in contrast to symmetric ears of $\lambda_c(z)$ in (b). Inset in (a) – ERV profile of a BatMR. (g) Relative nonuniformity $\varepsilon(z)$ of an eigenmode and constant eigenmode amplitude profile $\Delta\rho(z)$ at the BMR surface for different Q-factors of a silica BatMR with $n_r = 1.44$ at wavelength $\lambda_{mpq} = 1.55$ μm.

The original ERV of optical fibers is primarily caused by the frozen-in capillary waves having the order of an angstrom [12-14]. It can be found from Eq. (1) and rescaling equation $\Delta r_{eff}(z) = r_0 \Delta\lambda_{mp}^{(c)}(z)/\lambda_{mp}^{(c)}$ that the perturbation of the field along the uniform BatMR segment by the ERV spatial spectral component $\Delta r_k(z) = \Delta r_{0k} \exp(ikz)$ results in relative variation of the field magnitude

$$\varepsilon_k(z) = \Delta E_{mpq}^{(BMR)}(z,r_0,\varphi) / E_{mpq}^{(BMR)}(z,r_0,\varphi) = \varepsilon_{0k}\exp(ikz),$$
$$\varepsilon_{0k} = 8\pi^2 n_r^2 (\lambda_{mp}^{(c)}k)^{-2} r_0^{-1} \Delta r_{0k}. \quad \textbf{(6)}$$

For a silica fiber with characteristic $\Delta r_{0k} = 0.2$ nm, $r_0 = 50$ μm, $\lambda_{mp}^{(c)} = 1.5$ μm, and frozen-in wave spatial frequency $k = 1$ μm⁻¹, we find the relative field nonuniformity $\varepsilon_{0k} \cong 0.0003$.

Suppressing this ERV is currently challenging for lengthy optical fibers [14]. However, we suggest that it can be reduced to less than 1 Å by surface postprocessing developed in SNAP technology [1, 4-7].

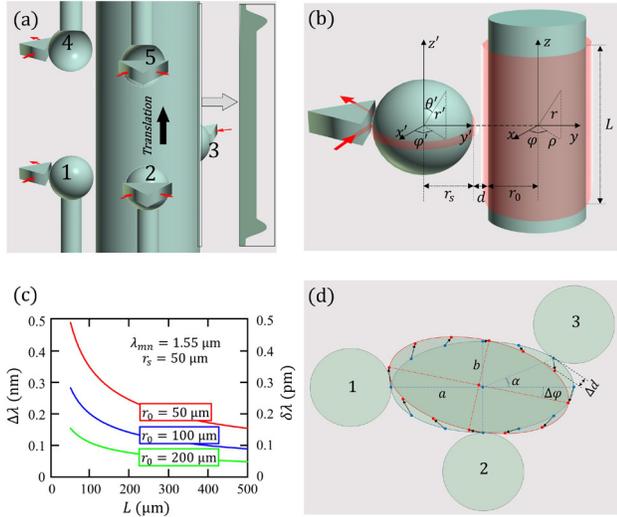

Fig. 2. (a) Translation of a BatMR coupled to five SMRs, 1,2,3,4, and 5. Inset – ERV profile of the BatMR. (b) SMR and a BatMR in coordinate systems. (c) Dependence of eigenwavelength splitting $\Delta\lambda$ and splitting variation $\delta\lambda$ caused by 1 Å change of BatMR-SMR separation $d = 200$ nm. (d) A BatMR with asymmetric cross-section coupled to three SMR before and after small rotation and displacement.

The unique uniformity of the bat mode amplitude suggests the application of a BatMR as an angstrom-precise *translation reference*. Besides fundamental interest, reaching the angstrom and eventually picometer precision of translation is critical for several applications in nanotechnology and nanoscience. In particular, solution of this problem is important in semiconductor manufacturing [15], atomic-scale electronic engineering [16, 17] as well as for manufacturing of metamaterial, plasmonic and nanophotonic devices [18]. Conventional approaches developed for ultraprecise linear translation are based on capacitive and piezo sensors and optical interferometers (see [19-24] and references therein). The stages with 10 pm resolution enabling translation over distances of 10 μm are available on the market [25, 26]. However, to our knowledge, the problem of translation with the subnanometer precise straightness and flatness has not been satisfactory explored. The reason is presumably in the absence of the reference which allows to follow the straight direction along the required length and with the required precision. The best optical flats, including those fabricated of silica, have the flatness of around 1 nm over the sub-millimeter areas [27], and their application to support the subnanometer-precise straight-line translation at microscale is problematic [28]. In contrast, we show below that a BatMR can be used as a reference for angstrom-precise straight and flat translation along its axial segment with length $L \sim 100$ μm.

The device proposed here consists of a BatMR and a set of spherical microresonators (SMRs), which are positioned in a submicron distance from the BatMR as illustrated in Fig. 2(a). Light is coupled into SMRs through prisms (shown in this figure) or fiber tapers with micron diameter waist [29]. The output light is detected by an optical spectrum analyzer not shown in Fig. 2(a). Each of SMR controls one degree of the BMR freedom by monitoring the splitting of resonance wavelengths [30, 31]. Therefore, in the absence of rotational BatMR symmetry (see below) we need five SMRs illustrated in Fig. 2(a) to fix all five BMR transverse degrees of freedom and enable its precise translation along a straight line.

We assume that each of the SMRs have a WGM $E^{(SMR)}_{m_s p_s q_s}(r', \theta', \varphi')$ with the wavelength eigenvalue $\lambda^{(SMR)}_{m_s p_s q_s}$ which has or tuned to have very small separation $\Delta\Lambda$ from bat mode eigenvalues $\lambda_{mpq}$ chosen different for different SMRs (it should have the same axial quantum number $q$ but different azimuthal quantum numbers $m$). A better accuracy of simultaneous matching of the eigenwavelengths of these bat modes and corresponding cutoff wavelengths of the fiber can be achieved for shallower and wider BatMR ears.

In close proximity of a SMR to the BatMR, the bat mode eigenvalue splits into two eigenvalues, which separation is determined as $(\Delta\lambda^2 + \Delta\Lambda^2)^{1/2}$ where [32]:

$$\Delta\lambda = \frac{1}{2n_r^2 k_0^2} \iint_S dx dz \left( (E^{(BMR)}_{mpq})^* \frac{dE^{(SMR)}_{m_s p_s q_s}}{dy} - (E^{(SMR)}_{m_s p_s q_s})^* \frac{dE^{(BMR)}_{mpq}}{dy} \right). \quad (7)$$

Here the wavenumber $k_0 = 2\pi/\lambda_{mpq}$, the BatRM and SMR modes are normalized, and the integral is taken along the plane $y = r_s + d/2$ where $r_s$ is the SMR radius and $d$ is the BatMR-SMR separation (Fig. 2(b)). The refractive index $n_r$ of all microresonators is assumed to be the same.

To enable the full control of the BatMR translation, below we explore a BatMR with an asymmetric cross-section. However, to estimate the SMR-BatMR coupling sensitivity, it is sufficient to use the axially symmetric model. We assume that the length $L$ of the BatMR segment with axially uniform $E^{(SMR)}_{mpq}(r, \theta, \varphi)$ is much greater than the length of bumps enabling this uniformity (see insert in Fig. 2(a)). Under the assumptions made, the integral in Eq. (7) is found analytically. As the result, the wavelength splitting for the SMR radial and axial quantum numbers $p_s, q_s = 0,1,2,3,4$ and BMR radial quantum numbers $p, q = 0,1,2,3,4$ is

$$\Delta\lambda = \frac{2\pi B_{p_s} C_{q_s} B_p \exp\left(-k_0 d\sqrt{n_r^2 - 1}\right)}{n_r^{1/4}(n_r^2 - 1)^{3/2} k_0^{11/4} r_s^{1/4} r_0^{1/2} (r_s + r_0)^{1/2} L^{1/2}}, \quad (8)$$

where $B = (2.018, 1.762, 1.634, 1.552, 1.494)$ and $C = (3.733, 0, 3.257, 0, 3.088)$. Due to the antisymmetric behavior of SMR modes on $z'$ for odd $p_s$, the splitting is zero for $p_s = 1$ and 3. The plots of $\Delta\lambda$ as a function of length $L$ for SMR radius $r_s = 50$ μm and BatMR radii $r_s = 50, 100$ and $200$ μm are shown in Fig. 2(c). The values of $\Delta\lambda$ are in reasonable agreement with the splitting between microspheres experimentally measured in Ref. [33]. In addition, Fig. 2(c) shows dependencies of the splitting variation $\delta\lambda$ corresponding to the change of 1 Å in SMR-BatMR separation $d = 200$ nm. These variations can be measured for microresonators with $Q \gtrsim 10^6$ and SMRs with precisely tuned eigenwavelengths to ensure $\Delta\Lambda \lesssim \Delta\lambda$. Indeed, then for $\lambda_{mn} = 1.55$ μm the FWHM of the resonance is $\gamma_{mn} \lesssim 1$ pm and variations $\delta\lambda$ can be measured with a power meter having smaller than 10% relative error. For precise measurement of $\delta\lambda$, the free spectral range along the axial quantum number $q$ of the WGM considered should be greater than or comparable with the resonance width $\gamma_{mp}$. This condition leads to the characteristic maximum length of translation $L_{max} = \frac{1}{8}\lambda_{mp} Q^{1/2} n_r^{-1}$. For example, for the same parameters as in Fig. 2, $L_{max} \cong 400$ μm for $Q = 10^7$. The splitting in Eq. (8) depends on the separation between a SMR and BatMR. Keeping it constant will force the BatMR to follow its surface profile.

The fundamental quantum limit of the measurement precision of the wavelength splitting $\Delta\lambda$ determined by Eq. (8) is [30, 31]

$$\delta\lambda_{quant} = \frac{1}{Q}\left(\frac{\hbar c}{\tau W \lambda_{mn}^{(c)}}\right)^{1/2}\left(\frac{\partial \Delta\lambda}{\partial d}\right)^{-1} = \frac{c^{1/2}\hbar^{1/2}(\lambda_{mn}^{(c)})^{3/2}}{2\pi(n_r^2-1)^{1/2}Q\Delta\lambda\tau^{1/2}W^{1/2}}. \quad (9)$$

Here $c$ is the speed of light $\hbar$ is the Plank constant, $W$ is the power of light in the optical spectrum analyzer used, $\Delta\lambda$ is determined from Eq. (8), and $\tau$ is the measurement time. Setting $Q = 10^7$, $\Delta\lambda = 0.1$ nm, $W = 10$ mW, and $\tau = 1$ ms, we find $\delta\lambda_{quant} \sim 10^{-4}$ pm, which is significantly smaller than the precision $\delta\lambda \sim 10^{-2}$ pm required for our measurements (Fig. 2(c)).

The major deviation from the straight-line translation is caused by the variation of BatMR local surface height and variation of SMRs and BatMR dimensions in time. While the angstrom-scale surface height variation of an actual BatMR can be recorded and taken into account in the process of translation, the temperature variation affects the measurement precision randomly. For the microresonator radii $r_0, r_s \sim 100$ μm and the temperature change $\lesssim 0.1°C$, we find their radius variation $\lesssim 0.1$Å, which insignificantly affects the angstrom-precise straight-line translation.

Finally, we determine the cross-sectional asymmetry of the BMR required to suppress its axial rotation in the process of translation. We assume that the BMR cross-section has the elliptic shape with semi-major and semi-minor axes $a$ and $b$ shown in Fig. 2(d). We position SMR 1 at the vertex and co-vertex of the ellipse and determine the position of SMR 3 to arrive at the maximum possible separation $\Delta d$ between it and BMR after the BMR is rotated by small angle $\Delta\varphi$. It is assumed that the separations BMR and SMR 1 and between BMR and SMR 2 is kept constant during this rotation. Cumbersome calculations yield the following simple result. The maximum separation $\Delta d = (a-b)\Delta\varphi$ is achieved for the SMR 3 located at angle $\alpha = \text{atan}[(b/a)^{3/2}]$ with respect to major axis of the elliptic cross-section (Fig. 2(d)). For the BMR with small asymmetry, $\alpha \cong 45^0$ and 1Å rotational displacement at the BMR surface corresponds to $\Delta\varphi \cong (1 \text{ Å})/a$. For example, assuming $a - b = 0.1a$ we find $\Delta d \cong 10$ pm. It follows from the above calculations that this small variation can be measured with microresonators having $Q \geq 10^6$ and relative measurement precision error smaller than 1%. For $Q \sim 10^7$ or for BMR with a greater rotational asymmetry, the required measurement precision can be relaxed to 10%. Thus, maintaining constant values of splitting $\Delta\lambda$ for all SMRs during translation allows us to use the constant-amplitude bat mode field as a translation reference enabling angstrom-precise straightness and flatness.

**Funding.** Engineering and Physical Sciences Research Council, (EP/P006183/1), Wolfson Foundation (22069).

**Disclosures**. The author declares no conflicts of interest.